\documentclass[11pt, oneside]{article}   	
\usepackage{geometry,amsmath,amssymb,amsthm,amsfonts,mathrsfs,marvosym,graphicx,graphics,keyval,young}
\usepackage{braket,ytableau}  

\usepackage[toc,page]{appendix}
\usepackage{graphicx}				
\usepackage{amssymb}

\newcommand{\mc}{\mathcal}
\newcommand{\mrn}{\mathrm}

\title{\bf{\huge{$SO(N)$ restricted Schur polynomials}}}
\author{\bf{\Large{Garreth Kemp$^{1}$\footnote{garreth.kemp@students.wits.ac.za}}}}
\date{}			
\begin{document}

\begin{titlepage}
\maketitle

\begin{center}
	\emph{$^{1}$National Institute for Theoretical Physics,\\
		Department of Physics and Centre for Theoretical Physics,\\
		University of the Witwatersrand, Wits, 2050,\\
		South Africa}
\end{center}

\begin{abstract}

We focus on the $1/4$-BPS sector of free super Yang-Mills theory with an $SO(N)$ gauge group. This theory has an AdS/CFT dual in the form of type IIB string theory with AdS$_{5}\times \mc{R}$P$^{5}$ geometry. With the aim of studying excited giant graviton dynamics, we construct an orthogonal basis for this sector of the gauge theory in this work. First, we demonstrate that the counting of states, as given by the partition function, and the counting of restricted Schur polynomials matches by restricting to a particular class of Young diagram labels. We then give an explicit construction of these gauge invariant operators and evaluate their two-point function exactly. This paves the way to studying the spectral problem of these operators and their $D$-brane duals.

\end{abstract}
\end{titlepage}
\tableofcontents

\section{Introduction}

The most studied example of AdS/CFT is $\mc{N}=4$ super Yang-Mills theory with $U(N)$ gauge group, and its dual, type IIB string theory on AdS$_{5}\times S^{5}$. Testing the AdS/CFT correspondence in non-planar limits of the gauge theory and its dual string theory is a very interesting problem. In this regard, integrability would be a powerful tool. Therefore, it is interesting to investigate if the integrability, that was found in the planar limit \cite{Beisert}, is also present in non-planar limits. Furthermore, testing AdS/CFT in less supersymmetric settings is also an interesting problem. In particular we have the $1/4$-BPS sector of the theory in mind. The study of restricted Schur polynomials has yielded progress in both these directions. A restricted Schur is a local operator in the gauge theory which can be built from a variety of fields. For instance, we can use the complex scalar Higgs fields, the fermion fields as well as the gauge fields to build these operators. We consider restricted Schurs built using two types of complex scalar fields, $Z$ and $Y$ say. In this case, the definition is

\begin{equation}
\label{eq:RSUNDef}
	\chi_{R(r,s)\alpha\beta}(Z,Y) = \frac{1}{n!m!}\sum\limits_{\sigma \in S_{n+m}} \mrn{Tr}\big( P_{R\rightarrow (r,s)\alpha\beta} \Gamma_{R}(\sigma) \big) \mrn{Tr}\big( \sigma Z^{\otimes n}\otimes Y^{\otimes m} \big)
\end{equation}
\\
In (\ref{eq:RSUNDef}), $R$ is a Young diagram with $n+m$ boxes, corresponding to an irreducible representation (irrep) of $S_{n+m}$ and $(r,s)$ are a pair of Young diagrams with $n$ and $m$ boxes respectively corresponding to an irrep of $S_{n}\times S_{m}$. $P_{R\rightarrow (r,s)\alpha\beta}$ is a projector in the labels $R$ and $(r,s)$ and an intertwiner in the labels $\alpha$ and $\beta$. These labels are the multiplicity labels with which irrep $(r,s)$ is subduced from $R$ when restricting from $S_{n+m}$ to $S_{n}\times S_{m}$. Further details of (\ref{eq:RSUNDef}) can be found in \cite{GGWSA1}, \cite{GGO} for example. There are many good reasons to study the operators defined in (\ref{eq:RSUNDef}) as we now explain. 

The counting of states given by the partition function of the $1/4$-BPS sector of the free $U(N)$ theory was shown to match the counting of restricted Schurs that can be defined \cite{Collins}. Essentially, the partition function was expressed in terms of Littlewood Richardson coefficients which count the number of restricted Schurs for a given set of Young diagram labels, $R,(r,s)$. This result was generalised to an arbitrary product of $U(N)$ gauge groups in \cite{Quiver}. 

We are interested in operators whose bare dimension grows parametrically with $N$. For these operators, the large $N$ limit of correlation functions is not captured by summing only planar diagrams \cite{GGO}. The usual $1/N^{2}$ factor suppressing non-planar diagrams is over-powered by combinatorial factors resulting from evaluating all possible Wick contractions. Indeed, when the number of matrix fields in the operator scales with $N$, we are forced to sum an infinite number of non-planar diagrams in the large $N$ limit. A solution to this problem was given in \cite{Jevicki} for the $1/2$-BPS case. In this important and influential work, it was observed that there is a one-to-one correspondence between the space of $1/2$-BPS representations and Young diagrams, i.e., Schur polynomials. Using representation theory of the symmetric and unitary group, the two-point function of Schur polynomials was computed \emph{exactly} in the free theory limit. \cite{EMC} then achieved the remarkable result of computing the two-point function of restricted Schurs (\ref{eq:RSUNDef}) exactly in the free theory. The operators diagonalised the two-point function and a very simple formula for the final result was given. Thus, the restricted Schurs provide an exactly orthogonal basis for the $1/4$-BPS sector of the free $U(N)$ gauge theory.

Apart from the interesting properties of (\ref{eq:RSUNDef}) as gauge theory objects, restricted Schurs also have an AdS/CFT dual. When the scaling dimension of operators grows with $N$, $\chi_{R(r,s)\alpha\beta}(Z,Y)$, with $R$ having long columns or rows ($O(N)$ boxes), has been argued to be dual to a system of excited giant gravitons \cite{Invasion}, \cite{GGO}. Concretely, such a system is realised as a system of giant gravitons with strings attached \cite{GGWSA1}. Thus, restricted Schurs are a useful tool for studying non-perturbative physics of both the gauge theory and its dual string theory.

In a similar approach, \cite{Kimura:2007wy} wrote down an orthogonal basis of gauge invariant operators using $Z$ and $Z^{\dagger}$ using Brauer algebras and their permutation sub-algebras in the free theory limit. \cite{Ram} then also studied $1/4$ and $1/8$-BPS operators at zero coupling, $g^{2}_{YM} = 0$. Also by making heavy use of representation theory, the multi-matrix multi-trace operators constructed here were argued to form a basis and to diagonalise their two-point function. When the operators in \cite{Ram} consist of two matrix fields $Z$ and $Y$ say, they were shown to be identical to the basis of restricted Schur polynomials \cite{Collins}. Progress has also been made in the counting and constructing of $1/4$ and $1/8$-BPS operators at weak coupling \cite{Weakcoup}.

Tremendous progress has already been achieved in the study of the anomalous dimension spectrum of restricted Schur polynomials; see \cite{GGO}, \cite{NPI}, \cite{Spring}, \cite{twoloop}, \cite{HigherLoopAnDim} for examples. Anomalous dimensions of the restricted Schurs are dual to excitation energies of the excited giant graviton system. One of the main results of these works is that the spectrum of anomalous dimensions reduces to the set of normal mode frequencies of a system of decoupled harmonic oscillators. The fact that we get decoupled oscillators is evidence of integrability in the non-planar sectors studied in these works.

Is there a similar story for the theory with an $SO(N)$ gauge group? In this work, we take the first steps toward answering this question. $\mc{N} = 4$ super Yang-Mills theory with an $SO(N)$ gauge group is dual to type IIB string theory with AdS$_{5}\times \mc{R}$P$^{5}$ geometry. At the non-planar level, the spectral problems of $U(N)$ and $SO(N)$ gauge theories are distinct \cite{SpecProbSON}. Thus, there is a good chance that a study of anomalous dimensions of large dimension operators will reveal new aspects of D-brane physics. The $1/2$-BPS case for free $SO(N)$ gauge theory was studied in \cite{SON1}, \cite{SON2} and \cite{Diaz} . The results of \cite{SON1} and \cite{SON2} include defining Schur polynomials in the square of the eigenvalues and computing their two-point functions exactly in the free theory. Here too, the Schur polynomial was shown to diagonalise the two-point function, a property that was argued to be independent of the gauge group \cite{Diaz}.

We show that restricted Schur polynomials provide an exactly orthogonal basis for the $1/4$-BPS sector of free super Yang-Mills theory with $SO(N)$ gauge group. In this work, we take $N$ to be even. First we show that the counting of states obtained from the partition function in this sector matches the number of restricted Schurs that can be defined. This is achieved by expressing the partition function in terms of the Littlewood Richardson coefficients. The counting comes out differently from the $1/4$-BPS $U(N)$ case as we will see. $SO(N)$ has two invariant tensors. They are $\delta_{ij}$ and $\epsilon_{i_{1}i_{2}\cdots i_{N}}$. We focus on operators constructed using $\delta_{ij}$. We construct restricted Schur polynomials and manage to compute the two-point function exactly in the free theory. The result is expressed in a relatively simple formula. In the next section, we present some background theory needed for our computations. 

\section{Projectors}

In this section, we briefly discuss some of the projectors and states that appear frequently in our calculations. All projectors we discuss here project from the carrier space of irrep $R$ of $S_{2q}$, $q=n+m$, onto the carrier space of an irrep of some subgroup.

\subsection{From $R$ onto $(r,s)$}

Firstly we discuss the projectors $P_{R\rightarrow (r,s)\alpha\beta}$ constructed in \cite{GGO}. They project from $R$ onto irrep $(r,s)$ of the subgroup $S_{2n}\times S_{2m}$. In the multiplicity labels, however, $P_{R\rightarrow (r,s)\alpha\beta}$ is an intertwiner \cite{RepTh}. We write these operators as

\begin{equation}
\label{eq:1}
	P_{R\rightarrow (r,s)\alpha\beta} = \sum\limits^{d_{r}\times d_{s}}_{l=1} \ket{R(r,s)\alpha;l}\!\!\bra{R(r,s)\beta;l}
\end{equation}
\\
where $\ket{R(r,s)\alpha;l}$ is a state in the carrier space of the $\alpha$-th copy of irrep $(r,s)$ and $d_{r}\times d_{s}$ is the dimension of $(r,s)$. In this basis, the matrix representation of $\gamma \in S_{2n}\times S_{2m}$ in irrep $R$ is block diagonal with the $S_{2n}\times S_{2m}$ irreps as the diagonal blocks. For example 

\begin{equation}
\label{eq:2}
	\Gamma_{R}(\gamma) = \left(\begin{array}{cccc}\Gamma_{(r',s')}(\gamma) & 0 & 0 & 0 \\0 & \Gamma_{(r'',s'')}(\gamma) & 0 & 0 \\0 & 0 & \Gamma_{(r,s)1}(\gamma) & 0 \\0 & 0 & 0 & \Gamma_{(r,s)2}(\gamma)\end{array}\right)
\end{equation}
\\
where we have shown two copies of irrep $(r,s)$. There are four projectors that can be defined for $(r,s)$. Acting on $\Gamma_{R}(\gamma)$ each give the following

\begin{eqnarray}
	P_{R\rightarrow(r,s)11}\Gamma_{R}(\gamma) &=& \left(\begin{array}{cccc}0 & 0 & 0 & 0 \\0 & 0 & 0 & 0 \\0 & 0 & \Gamma_{(r,s)1}(\gamma) & 0 \\0 & 0 & 0 & 0\end{array}\right)\\
	P_{R\rightarrow(r,s)22}\Gamma_{R}(\gamma) &=& \left(\begin{array}{cccc}0 & 0 & 0 & 0 \\0 & 0 & 0 & 0 \\0 & 0 & 0 & 0 \\0 & 0 & 0 & \Gamma_{(r,s)2}(\gamma)\end{array}\right)\\
	P_{R\rightarrow(r,s)12}\Gamma_{R}(\gamma) &=& \left(\begin{array}{cccc}0 & 0 & 0 & 0 \\0 & 0 & 0 & 0 \\0 & 0 & 0 & \Gamma_{(r,s)2}(\gamma) \\0 & 0 & 0 & 0\end{array}\right)\\
	P_{R\rightarrow(r,s)21}\Gamma_{R}(\gamma) &=& \left(\begin{array}{cccc}0 & 0 & 0 & 0 \\0 & 0 & 0 & 0 \\0 & 0 & 0 & 0 \\0 & 0 & \Gamma_{(r,s)1}(\gamma) & 0\end{array}\right)
\end{eqnarray}
\\
\subsection{From $R$ onto $[S]$}

Let $R$ have an even number of boxes in each row. Restricting $S_{2q}$ to the subgroup $S_{q}[S_{2}]$, one can find a basis in which $\Gamma_{R}(\xi)$, $\xi\in S_{q}[S_{2}]$, is block diagonal with the $S_{q}[S_{2}]$ irreps appearing on the diagonal. For $R$ having even rows, there exists a 1-dimensional irrep of $S_{q}[S_{2}]$ which we denote $[S]$. In this irrep, all matrices are represented by 1. Let $\ket{[S]}$ be the state spanning the carrier space of $[S]$. Then we have \cite{Ivanov}

\begin{equation}
\label{eq:3}
	\Gamma_{R}(\xi)\ket{[S]} = \ket{[S]}, \hspace{20pt} \xi \in S_{q}[S_{2}].
\end{equation}
\\
We can also define a projector to go from $R$ to $[S]$.

\begin{equation}
\label{eq:4}
	P_{[S]} = \frac{1}{q!2^{q}}\sum\limits_{\xi\in S_{q}[S_{2}]}\Gamma_{R}(\xi)
\end{equation}
\\
The state $\ket{[S]}$ may be calculated as the eigenvector of $P_{[S]}$ with eigenvalue 1. Lastly, $[S]$ is subduced from $R$ with multiplicity 1 \cite{SON1}.

\subsection{From $(r,s)\beta$ to $([A],[A])\beta$}

Recall that $(r,s)\beta$ is the $\beta$-th copy of the irrep of $S_{2n}\times S_{2m}$. If $r$ has an even number of boxes in each column, then, upon restricting to the $S_{n}[S_{2}]$ subgroup another type of 1 dimensional irrep may be subduced \cite{Ivanov}. Call this irrep $[A]$. In this irrep, 

\begin{equation}
\label{eq:5}
	\Gamma_{r}(\eta) = \mrn{sgn}(\eta), \hspace{20pt} \eta \in S_{n}[S_{2}].
\end{equation}
\\
Denote by $\ket{[A]}$ the state spanning the 1-dimensional carrier space of $[A]$. Irrep $[A]$ is also subduced from $r$ with no multiplicity \cite{SON1}. If $r$ and $s$ both have even columns, then each may subduce the irrep $[A]$ of $S_{n}[S_{2}]$ and $S_{m}[S_{2}]$ respectively. Thus, $(r,s)\beta$ subduces the irrep $([A],[A])\beta$ of $S_{n}[S_{2}]\times S_{m}[S_{2}]$. Denote by $\ket{[A],[A]\beta}$ the state spanning this 1-dimensional irrep. Define

\begin{equation}
\label{eq:6}
	P_{R\rightarrow ([A],[A])\beta} = P_{R\rightarrow(r,s)\beta\beta}P_{[A,A]}, \hspace{20pt} P_{[A,A]} = \frac{1}{n!m!2^{q}}\sum\limits_{\eta \in S_{n}[S_{2}]\times S_{m}[S_{2}]}\mrn{sgn}(\eta)\Gamma_{R}(\eta)
\end{equation}
\\
The state $\ket{[A],[A]\beta}$ may be defined to be the eigenvector of $P_{R\rightarrow ([A],[A])\beta}$ with eigenvalue 1. This state has the following properties

\begin{enumerate}
  \item \begin{equation}
\label{eq:611}
	P_{R\rightarrow (r,s)\alpha\beta} \ket{[A],[A]\beta} = \ket{[A],[A]\alpha}
\end{equation}
  \item 
  \begin{equation}
  	\Gamma_{R}(\eta) \ket{[A],[A]\beta} = \mrn{sgn}(\eta)\ket{[A],[A]\beta},\hspace{20pt} \eta \in S_{n}[S_{2}]\times S_{m}[S_{2}]
	\end{equation}
	\item The quantity $\langle[A],[A]\beta | R(r,s)\beta;l\rangle $ does not depend on the multiplicity index $\beta$. This will be important when we construct the restricted Schurs. 
\end{enumerate}

\section{Counting}

In this section we express the partition function for the $1/4$-BPS sector of free $SO(2n)$ gauge theory in terms of the Littlewood Richardson coefficients depending on three Young diagram labels. Two Young diagrams will have an even number of boxes in each \emph{column}, and the third one (induced from the previous two) will have an even number of boxes in each \emph{row}. This is first demonstrated for the $SO(4)$ case for simplicity. It is then easy to extend the argument to the general $SO(2n)$ case, which follows thereafter. The Littlewood Richardson number $g(\lambda,\mu,\xi)$ counts the number of times a Young diagram $\xi$ is induced from two smaller Young diagrams $\mu$ and $\lambda$, say. Thus, $g(\lambda,\mu,\xi)$ counts the number of restricted Schurs that may be defined for labels $\{\lambda,\mu,\xi\}$. 

The partition function for the $1/4$-BPS sector of free $SO(2n)$ gauge theory is \cite{SON1}, \cite{Dolan1}

\begin{eqnarray}
\label{eq:SO2nPF}
	G(t_{1},t_{2}) &=& \frac{1}{2^{n-1}n!}\int_{T_{n}} \prod\limits^{n}_{j=1} \frac{dx_{j}}{2\pi ix_{j} }\Delta(x+x^{-1})^{2}\times  \\
	&& \prod\limits^{2}_{k=1}\prod\limits_{1\leqslant i<j\leqslant n}\frac{1}{(1-t_{k})^{n}}\frac{1}{(1-t_{k}x_{i}x_{j})(1-t_{k}x_{i}x^{-1}_{j})(1-t_{k}x^{-1}_{i}x_{j})(1-t_{k}x^{-1}_{i}x^{-1}_{j})} \nonumber
\end{eqnarray}
\\
Let's study the simple $SO(4)$ case first and generalise to $SO(2n)$ thereafter. For $SO(4)$, this becomes

\begin{eqnarray}
\label{eq:1steqnforG}
	G(t_{1},t_{2}) &=& \frac{1}{4}\int \frac{dx_{1}}{2\pi ix_{1}}\int \frac{dx_{2}}{2\pi ix_{2}}\Delta(x+x^{-1})^{2}\times  \\
	&& \prod\limits^{2}_{k=1}\frac{1}{(1-t_{k})^{2}}\frac{1}{(1-t_{k}x_{1}x_{2})(1-t_{k}x_{1}x^{-1}_{2})(1-t_{k}x^{-1}_{1}x_{2})(1-t_{k}x^{-1}_{1}x^{-1}_{2})}\nonumber
\end{eqnarray}
\\
The factor in the second line may be written as

\begin{eqnarray}
	&&\frac{1}{(1-t_{k}x_{1}x^{-1}_{1})(1-t_{k}x_{2}x^{-1}_{2})(1-t_{k}x_{1}x_{2})(1-t_{k}x_{1}x^{-1}_{2})(1-t_{k}x^{-1}_{1}x_{2})(1-t_{k}x^{-1}_{1}x^{-1}_{2})} \nonumber \\ 
\label{eq:rewritingprod}
	&&\hspace{200pt} =\; \prod\limits_{1\leqslant i< j \leqslant 4}\frac{1}{1-t_{k}y_{i}y_{j}}
\end{eqnarray}
\\
with $y_{1}=x_{1}, y_{2} = x^{-1}_{1}, y_{3} = x_{2}, y_{4} = x^{-1}_{2}$. Now we may expand this product in terms of Schur polynomials using the formula \cite{Ellitic}

\begin{equation}
\label{eq:modcaulittd}
	\prod\limits_{1\leqslant i< j \leqslant L}\frac{1}{1-y_{i}y_{j}} = \sum\limits_{\mu}s_{\mu^{2}}(y_{1},..., y_{L})
\end{equation}
\\
where $\mu^{2}$ is defined as the partition with parts

\begin{equation}
\label{eq:7}
	(\mu^{2})_{i} = \mu_{\lceil i/2 \rceil}.
\end{equation}
\\
This means the following. Take, for example, $i = \{1,2,3,4\}$. Then, $i/2 = \{ 1/2, 1, 3/2, 2 \}$. The ceiling symbol then tells us to take: $\lceil i/2 \rceil = \{ 1,1,2,2 \}$. So, if $\mu^{2}\,_{i}$ denotes the $i$th part of $\mu^{2}$, then 

\begin{equation}
\label{eq:8}
	\mu^{2} = (\mu^{2}\,_{1},\mu^{2}\,_{2},\mu^{2}\,_{3},\mu^{2}\,_{4}) = (\mu_{1},\mu_{1},\mu_{2},\mu_{2})
\end{equation}
\\
Thus, in general for a partition $\mu^{2}$ with $n$ parts,

\begin{equation}
\label{eq:9}
	\mu^{2} = (\mu^{2}\,_{1},\mu^{2}\,_{2},\mu^{2}\,_{3},\mu^{2}\,_{4},\cdots, \mu^{2}\,_{n-1},\mu^{2}\,_{n}) = (\mu_{1},\mu_{1},\mu_{2},\mu_{2},\cdots,\mu_{n/2},\mu_{n/2})
\end{equation}
\\
Partitions $\mu^{2}$ then correspond to Young diagrams with an even number of boxes in each column. Including the prefactor $t_{k}$ in (\ref{eq:modcaulittd}), we get

\begin{equation}
\label{eq:911}
	 \prod\limits_{1\leqslant i< j \leqslant L}\frac{1}{1-t_{k}y_{i}y_{j}} = \sum\limits_{\mu}s_{\mu^{2}}(\sqrt{t_{k}}y_{1},..., \sqrt{t_{k}}y_{L}) = \sum\limits_{\mu} (t_{k})^{|\mu^{2}|/2} s_{\mu^{2}}(y_{1},...,y_{L})
\end{equation}
\\
where we used the fact that Schur polynomial $s_{\lambda}(y_{1},y_{2},\cdots, y_{L})$ may be written as 

\begin{equation}
\label{eq:10}
	s_{\lambda}(ty_{1},ty_{2},\cdots,ty_{L}) = t^{|\lambda|}s_{\lambda}(y_{1},y_{2},\cdots, y_{L}).
\end{equation}
\\
In the above formulas, $|\,|$ stands for the weight of the partition. The partition function (\ref{eq:1steqnforG}) becomes

\begin{eqnarray}
\label{eq:11}
	G(t_{1},t_{2}) &=& \frac{1}{4}\int \frac{dx_{1}}{2\pi ix_{1}}\int \frac{dx_{2}}{2\pi ix_{2}}\Delta(x+x^{-1})^{2}\times  \\
	&&\sum\limits_{\lambda}\sum\limits_{\mu}(t_{1})^{|\lambda^{2}|/2}(t_{2})^{|\mu^{2}|/2}s_{\lambda^{2}}(x_{1},x^{-1}_{1},x_{2},x^{-1}_{2})s_{\mu^{2}}(x_{1},x^{-1}_{1},x_{2},x^{-1}_{2})\nonumber
\end{eqnarray}
\\
Now, we use the product rule involving Schur polynomials. This means that we can write the product of two Schur polynomials with partitions $\lambda^{2}$ and $\mu^{2}$ as the sum over a single Schur with partition $\xi$ times by the Littlewood Richardson coefficient $g(\lambda^{2},\mu^{2},\xi)$ \cite{Dolan2}. The Littlewood Richardson coefficient is how many times $\xi$ subduces $(\lambda^{2},\mu^{2})$. Thus,

\begin{eqnarray}
\label{eq:12}
	G(t_{1},t_{2}) &=& \sum\limits_{\xi}\sum\limits_{\lambda}\sum\limits_{\mu} g(\lambda^{2},\mu^{2},\xi) (t_{1})^{|\lambda^{2}|/2}(t_{2})^{|\mu^{2}|/2} \times\\
	&&  \frac{1}{4}\int \frac{dx_{1}}{2\pi ix_{1}}\int \frac{dx_{2}}{2\pi ix_{2}}\Delta(x+x^{-1})^{2}s_{\xi}(x_{1},x^{-1}_{1},x_{2},x^{-1}_{2}) \nonumber
\end{eqnarray}
\\
Next, we generalise this to $SO(2n)$. In the same way as in equation (\ref{eq:rewritingprod})





\begin{equation}
	\label{eq:13}
	\frac{1}{(1-t_{k})^{n}}\prod\limits_{1\leqslant i <j \leqslant n}\frac{1}{(1-t_{k}x_{i}x_{j})(1-t_{k}x_{i}x^{-1}_{j})(1-t_{k}x^{-1}_{i}x_{j})(1-t_{k}x^{-1}_{i}x^{-1}_{j})} = \prod\limits_{1\leqslant i< j \leqslant N}\frac{1}{1-t_{k}y_{i}y_{j}}
\end{equation}
\\
with $y_{i} = x_{\lceil i/2 \rceil}$ and $y_{2i} = x^{-1}_{i}$ for $i = 1,2,\cdots, n$. Using equation (\ref{eq:modcaulittd}) for each factor,

\begin{equation}
\label{eq:14}
	\prod\limits^{2}_{k=1} \prod\limits_{1\leqslant i< j \leqslant N}\frac{1}{1-t_{k}y_{i}y_{j}} = \sum\limits_{\lambda}\sum\limits_{\mu}(t_{1})^{|\lambda^{2}|/2}(t_{2})^{|\mu^{2}|/2}s_{\lambda^{2}}(y_{1},..,y_{N})s_{\mu^{2}}(y_{1},..,y_{N})
\end{equation}
\\
After using the product rule again, the $SO(2n)$ partition function in (\ref{eq:SO2nPF}) becomes

\begin{eqnarray}
\label{eq:InteformofPF}
	G(t_{1},t_{2}) &=& \sum\limits_{\xi}\sum\limits_{\lambda}\sum\limits_{\mu} g(\lambda^{2},\mu^{2},\xi) (t_{1})^{|\lambda^{2}|/2}(t_{2})^{|\mu^{2}|/2} \times\\
	&&  \frac{1}{2^{n-1}n!}\int_{T_{n}} \prod\limits^{n}_{j=1} \frac{dx_{j}}{2\pi ix_{j}}\Delta(x+x^{-1})^{2}s_{\xi}(x_{1},x^{-1}_{1},...,x_{n},x^{-1}_{n})\nonumber
\end{eqnarray}
\\
Recognising the Haar (or $G$-invariant) measure for $SO(2n)$, the integral in (\ref{eq:InteformofPF}) may be written as

\begin{equation}
\label{eq:15}
		I_{n} = \int_{SO(2n)}[dO] \,s_{\xi}(\mrn{x})
\end{equation}
\\
This integral is equal to 1 for two cases. The first is if $\xi$ has an even number of boxes in each \emph{row}. The second is if $\xi$ has a Young diagram with an even number of boxes in each row stuck onto a single column of $2n$ boxes\footnote{Recall that there are only two invariant tensors for $SO(2n)$. They are $\delta_{ij}$ and $\epsilon_{i_{1}i_{2}\cdots i_{2n}}$. This case is relevent for operators that are built using $\epsilon_{i_{1}i_{2}\cdots i_{2n}}$.}. Here are examples of each for $SO(4)$

\begin{equation}
\label{eq:16}
	\xi = \begin{Young} &&&\cr&\cr&\cr \end{Young}, \hspace{20pt} \mrn{or} \hspace{20pt} \xi = \begin{Young} &&\cr&&\cr\cr\cr \end{Young}
\end{equation}
\\
The integral vanishes in all other cases. See \cite{Ellitic} \cite{Vanishing} and \cite{SymmFunc}. We have checked these results for many examples for $SO(4)$ and $SO(6)$. With this result, the $SO(2n)$ partition function becomes

\begin{equation}
\label{eq:FinalPF}
	G(t_{1},t_{2}) = \sum\limits_{\xi}\sum\limits_{\lambda}\sum\limits_{\mu} g(\lambda^{2},\mu^{2},\xi) (t_{1})^{|\lambda^{2}|/2}(t_{2})^{|\mu^{2}|/2} 
\end{equation}
\\
where $\xi$ is a Young diagram with even rows and $\lambda^{2}$ and $\mu^{2}$ are Young diagrams with even columns. This result is truly different from the $U(N)$ $1/4$-BPS case, studied in \cite{Collins}. For $U(N)$ the counting went according to the square of the Littlewood-Richardson number. Here, the counting goes according to the Littlewood Richardson number itself. In other words, for a given $S_{2n}\times S_{2m}$ irrep, $(r,s)$, each with even columns, and given $S_{2q}$ irrep $R$ with even rows, the number of restricted Schurs one can define is exactly $g(r,s,R)$. We present some counting examples in appendix 1.

\section{Defining the restricted Schurs}

We now discuss the problem of defining the restricted Schurs for the $1/4$-BPS free $SO(N)$ gauge theory. We focus on operators that can be constructed using $\delta$ rather than $\epsilon_{i_{1}i_{2}\cdots i_{2n}}$. We construct these operators out of two complex matrix fields $Z$ and $Y$. The two-point function of these fields is given by \cite{SON1}

\begin{equation}
\label{eq:17}
	\langle Z^{ij}\overline{Z}^{kl} \rangle = \langle Z^{ij}Z_{kl} \rangle = \delta^{i}_{k}\delta^{j}_{l} - \delta^{i}_{l}\delta^{j}_{k}
\end{equation}
\\
and similarly for $Y$. We define our $SO(N)$ restricted Schurs to be

\begin{equation}
\label{eq:RSDef}
	O_{R(r,s)\alpha}(Z,Y) = \frac{1}{(2n)!(2m)!}\sum\limits_{\sigma \in S_{2q}} \mrn{Tr}\big( \mc{O}_{R(r,s)\alpha}\Gamma_{R}(\sigma)\big)C^{4\nu}_{I}\sigma^{I}_{J} (Z^{\otimes 2n} \otimes Y^{\otimes 2m})^{J}
\end{equation}
\\
where 

\begin{equation}
\label{eq:ODef}
	 \mc{O}_{R(r,s)\alpha} = \ket{[S]}\!\!\bra{[A],[A]\beta}P_{R\rightarrow (r,s)\beta\alpha}
\end{equation}
\\
In (\ref{eq:RSDef}), the tensor $(Z^{\otimes 2n} \otimes Y^{\otimes 2m})^{J}$ is defined as

\begin{equation}
\label{eq:18}
	(Z^{\otimes 2n} \otimes Y^{\otimes 2m})^{J} = Z^{j_{1}j_{2}}\cdots Z^{j_{2n-1}j_{2n}}Y^{j_{2n+1}j_{2n+2}}\cdots Y^{j_{2q-1}j_{2q}}
\end{equation}
\\
$C^{4\nu}_{I}$ is the contractor defined in \cite{SON2}

\begin{equation}
\label{eq:19}
	C^{4\nu}_{I} = \delta_{k_{1}k_{2}}\cdots \delta_{k_{2q-1}k_{2q}}(\sigma_{4\nu})^{K}_{I}
\end{equation}
\\
$C^{4\nu}_{I}$ is responsible for contracting the free indices in such a way to make the operator gauge invariant. For $\sigma_{4\nu}$, we take

\begin{equation}
\label{eq:20}
	\sigma_{4\nu} =  (1,2,3,4)(5,6,7,8)\cdots (2q-3,2q-2,2q-1,2q)
\end{equation} 
\\
This permutation contracts indices $1\&4$, $2\&3$ $\cdots$, $2q-3\&2q$ and $2q-2\&2q-1$, i.e.,

\begin{equation}
\label{eq:21}
	C^{4\nu}_{I}(\sigma)^{I}_{J} = \sigma^{i_{1}i_{2}i_{2}i_{1}\cdots i_{q-1}i_{q}i_{q}i_{q-1}}_{j_{1}j_{2}j_{3}j_{4}\cdots j_{2q}}.
\end{equation}
\\
$R$ is an irrep of $S_{2q}$ and $(r,s)$ is an irrep of the subgroup $S_{2n}\times S_{2m}$. Indices $\alpha$ and $\beta$ are multiplicity labels for the irrep $(r,s)$. State $\ket{[A],[A]\beta}$ is the state spanning the one-dimensional carrier space $([A],[A])\beta$ of $S_{n}[S_{2}]\times S_{m}[S_{2}]$ as defined in section 2. Concretely, it is the eigenvector of the operator $P_{R\rightarrow ([A],[A])\beta}$ with eigenvalue 1. State $\ket{[S]}$ is the state spanning the 1-dimensional carrier space $[S]$ of $S_{q}[S_{2}]$ as defined in section 2. The $S_{q}[S_{2}]$ here is chosen to stabilise the set $(1,4),(2,3),\cdots, (2q-3,2q),(2q-2,2q-1)$. Concretely, $\ket{[S]}$ is the eigenvector of $P_{[S]}$.

The coefficients in (\ref{eq:RSDef}) satisfy the following property

\begin{equation}
\label{eq:22}
	\mrn{Tr}\big( \mc{O}_{R(r,s)\alpha} \Gamma_{R}(\sigma) \big)\mrn{sgn}(\eta) = \mrn{Tr}\big( \mc{O}_{R(r,s)\alpha} \Gamma_{R}(\eta\sigma\xi) \big), \hspace{20pt} \xi \in S_{q}[S_{2}], \eta \in S_{n}[S_{2}]\times S_{m}[S_{2}]
\end{equation}
\\
Lastly, the operators in (\ref{eq:RSDef}) depend only on a single multiplicity label and thus, match the counting found in section 3.


\section{Two-point function}

In this section, we evaluate the two-point function of the operators defined in section 4. Recall the definition for our restricted Schurs,

\begin{eqnarray}
	O_{R(r,s)\alpha} &=& \frac{1}{(2n)!(2m)!}\sum\limits_{\sigma\in S_{2q}}\mrn{Tr}(\mc{O}_{R(r,s)\alpha}\Gamma_{R})C^{4\nu}_{I}\sigma^{I}_{J}(Z^{\otimes 2n} \otimes Y^{\otimes 2m})^{J}\\
	\overline{O}_{T(t,u)\beta} &=& \frac{1}{(2n)!(2m)!}\sum\limits_{\sigma\in S_{2q}}\mrn{Tr}(\mc{O}_{T(t,u)\beta}\Gamma_{R})C^{K}_{4\nu}(\overline{\sigma})^{L}_{K}(Z^{\otimes 2n} \otimes Y^{\otimes 2m})_{L}
\end{eqnarray}
\\
where

\begin{equation}
\label{eq:23}
	(\overline{\sigma})^{J}_{I} = \delta^{j_{\sigma(1)}}_{i_{1}}\delta^{j_{\sigma(2)}}_{i_{2}}\cdots \delta^{j_{\sigma(2q)}}_{i_{2q}} = \delta^{j_{1}}_{i_{\sigma^{-1}(1)}}\delta^{j_{2}}_{i_{\sigma^{-1}(2)}}\cdots \delta^{j_{2q}}_{i_{\sigma^{-1}(2q)}} = (\sigma^{-1})^{J}_{I}.
\end{equation}
\\
The first step in the calculation is evaluating the correlator 

\begin{equation}
\label{eq:CorrZY}
	\langle (Z^{\otimes 2n} \otimes Y^{\otimes 2m})^{J}(Z^{\otimes 2n} \otimes Y^{\otimes 2m})_{L} \rangle
\end{equation}
\\
In \cite{SON1} the correlator $(\langle Z^{\otimes 2n})^{J}(Z^{\otimes 2n})_{L} \rangle$ was found to be a sum over permutations belonging to the wreath product $S_{n}[S_{2}]$, where each term in the sum was weighted by the $\mrn{sgn}$ of the permutation. The correlator in (\ref{eq:CorrZY}) generalises to a sum over the subgroup $S_{n}[S_{2}]\times S_{m}[S_{2}]$

\begin{equation}
\label{CorrZYanswer}
	\langle (Z^{\otimes 2n} \otimes Y^{\otimes 2m})^{J}(Z^{\otimes 2n} \otimes Y^{\otimes 2m})_{L} = \sum\limits_{\eta \in S_{n}[S_{2}]\times S_{m}[S_{2}]} \mrn{sgn}(\eta) \eta^{J}_{L}
\end{equation}
\\
This formula may be easily checked for the example of $n=m=2$. 

\begin{eqnarray}
	&&\langle Z^{i_{1}i_{2}}Z^{i_{3}i_{4}}Y^{i_{5}i_{6}}Y^{i_{7}i_{8}}\bar{Z}^{j_{1}j_{2}}\bar{Z}^{j_{3}j_{4}}\bar{Y}^{j_{5}j_{6}}\bar{Y}^{j_{7}j_{8}} \rangle = \nonumber \\
	&&\langle Z^{i_{1}i_{2}}\bar{Z}^{j_{1}j_{2}} \rangle\langle  Z^{i_{3}i_{4}}\bar{Z}^{j_{3}j_{3}}  \rangle \Big[ \langle Y^{i_{5}i_{6}}\bar{Y}^{j_{5}j_{6}} \rangle\langle  Y^{i_{7}i_{8}}\bar{Y}^{j_{7}j_{8}}  \rangle +  \langle Y^{i_{5}i_{6}}\bar{Y}^{j_{7}j_{8}} \rangle\langle  Y^{i_{7}i_{8}}\bar{Y}^{j_{5}j_{6}}  \rangle  \Big] +\nonumber\\
	&& \langle Z^{i_{1}i_{2}}\bar{Z}^{j_{3}j_{4}} \rangle\langle  Z^{i_{3}i_{4}}\bar{Z}^{j_{1}j_{2}}  \rangle \Big[ \langle Y^{i_{5}i_{6}}\bar{Y}^{j_{5}j_{6}} \rangle\langle  Y^{i_{7}i_{8}}\bar{Y}^{j_{7}j_{8}}  \rangle +  \langle Y^{i_{5}i_{6}}\bar{Y}^{j_{7}j_{8}} \rangle\langle  Y^{i_{7}i_{8}}\bar{Y}^{j_{5}j_{6}}  \rangle  \Big]\nonumber\\
	&=& \langle Z^{i_{1}i_{2}}\bar{Z}^{j_{1}j_{2}} \rangle\langle  Z^{i_{3}i_{4}}\bar{Z}^{j_{3}j_{3}}  \rangle  \langle Y^{i_{5}i_{6}}Y^{i_{7}i_{8}}\bar{Y}^{j_{5}j_{6}}\bar{Y}^{j_{7}j_{8}} \rangle +  \langle Z^{i_{1}i_{2}}\bar{Z}^{j_{3}j_{4}} \rangle\langle  Z^{i_{3}i_{4}}\bar{Z}^{j_{1}j_{2}}  \rangle  \langle Y^{i_{5}i_{6}}Y^{i_{7}i_{8}}\bar{Y}^{j_{5}j_{6}}\bar{Y}^{j_{7}j_{8}} \rangle \nonumber\\
	&=&  \langle Z^{i_{1}i_{2}}Z^{i_{3}i_{4}}\bar{Z}^{j_{1}j_{2}}\bar{Z}^{j_{3}j_{4}} \rangle  \langle Y^{i_{5}i_{6}}Y^{i_{7}i_{8}}\bar{Y}^{j_{5}j_{6}}\bar{Y}^{j_{7}j_{8}} \rangle \nonumber\\
	&=& \bigg[\sum\limits_{\rho\in S_{2}[S_{2}]}\mrn{Sgn}(\rho)\rho\bigg] \bigg[\sum\limits_{\sigma\in S_{2}[S_{2}]}\mrn{Sgn}(\sigma)\sigma\bigg]\nonumber\\
	&=& \sum\limits_{\psi \in S_{2}[S_{2}]\times S_{2}[S_{2}]}\mrn{Sgn}(\psi)\psi^{I}_{J}
\end{eqnarray}
\\
where $\psi^{I}_{J} = \rho \sigma$ and $\mrn{Sgn}(\psi)=\mrn{Sgn}(\rho)\mrn{Sgn}(\sigma)$. Now compute the two-point function

\begin{eqnarray}
	\langle O_{R(r,s)\alpha}\overline{O}_{T(t,u)\beta} \rangle &=& \frac{1}{\big( (2n)!(2m)! \big)^{2}}\sum\limits_{\sigma,\rho\in S_{2q}}\mrn{Tr}\big(\mc{O}_{R(r,s)\alpha}\Gamma_{R}(\sigma)\big)\mrn{Tr}\big(\mc{O}_{T(t,u)\beta}\Gamma_{T}(\rho)\big)C^{4\nu}_{I}C^{L}_{4\nu}\sigma^{I}_{J}(\rho^{-1})^{K}_{L} \nonumber \\
	&& \times \langle (ZY)^{J}(ZY)_{K} \rangle \nonumber \\
	&=& \frac{1}{\big( (2n)!(2m)! \big)^{2}}\sum\limits_{\sigma,\rho\in S_{2q}}\mrn{Tr}\big(\mc{O}_{R(r,s)\alpha}\Gamma_{R}(\sigma)\big)\mrn{Tr}\big(\mc{O}_{T(t,u)\beta}\Gamma_{T}(\rho)\big)C^{4\nu}_{I}C^{L}_{4\nu}\sigma^{I}_{J}(\rho^{-1})^{K}_{L} \nonumber \\
	&& \sum\limits_{\eta \in S_{n}[S_{2}]\times S_{m}[S_{2}]}\mrn{sgn}(\eta)\eta^{J}_{K} \nonumber \\
	&=& \frac{1}{\big( (2n)!(2m)! \big)^{2}}\sum\limits_{\sigma,\rho\in S_{2q}}\sum\limits_{\eta \in S_{n}[S_{2}]\times S_{m}[S_{2}]}\mrn{Tr}\big(\mc{O}_{R(r,s)\alpha}\Gamma_{R}(\sigma)\big)\mrn{Tr}\big(\mc{O}_{T(t,u)\beta}\Gamma_{T}(\rho)\big)\mrn{sgn}(\eta) \nonumber \\
\label{eq:TPF1}
	&& \hspace{50pt} \times \;C^{4\nu}_{I}C^{L}_{4\nu}(\rho^{-1}\eta\sigma)^{I}_{L}
\end{eqnarray}
\\
Relabelling the sum over $\sigma = \eta^{-1}\psi$ where $\psi \in S_{2q}$, we obtain 
\\
\begin{equation}
\label{eq:TPF2}
	\langle O_{R(r,s)\alpha}\overline{O}_{T(t,u)\beta} \rangle  = \frac{n!m!2^{q}}{\big( (2n)!(2m)! \big)^{2}} \sum\limits_{\psi,\rho\in S_{2q}}\mrn{Tr}\big(\mc{O}_{R(r,s)\alpha}\Gamma_{R}(\psi)\big)\mrn{Tr}(\mc{O}_{T(t,u)\beta}\Gamma_{T}(\rho))C^{4\nu}_{I}C^{L}_{4\nu}(\rho^{-1}\psi)^{I}_{L}
\end{equation}
\\
We also used the fact that $\eta \in S_{n}[S_{2}]\times S_{m}[S_{2}]$, which is a subgroup of $S_{2n}\times S_{2m}$ and that

\begin{equation}
\label{eq:24}
	\bra{[A],[A]\gamma}\Gamma_{R}(\eta^{-1}) = \bra{[A],[A]\gamma}\mrn{sgn}(\eta^{-1})
\end{equation}
\\
which then canceled with $\mrn{sgn}(\eta)$ in (\ref{eq:TPF1}). Next, relabel the sum over $\psi$ by letting $ \psi = \rho\tau$, with $\tau \in S_{2q}$. After using the orthogonality relation

\begin{equation}
\label{eq:25}
	\sum\limits_{\rho \in S_{2q}} \Gamma_{R}(\rho)_{ij}\Gamma_{T}(\rho)_{kl} = \frac{(2q)!}{d_{R}}\delta_{RT}\delta_{ik}\delta_{jl}
\end{equation}
\\
and using

\begin{equation}
\label{eq:26}
	\mc{O}_{R(r,s)\alpha}\mc{O}^{T}_{T(t,u)\beta} = \ket{[S]}\!\!\bra{[A],[A]\gamma}P_{R\rightarrow(r,s)\gamma\zeta} \ket{[A],[A]\zeta}\!\!\bra{[S]}\delta_{RT}\delta_{rt}\delta_{su}\delta_{\alpha\beta},
\end{equation}
\\
where $T$ in the superscript is the transpose of $\mc{O}$, equation (\ref{eq:TPF2}) becomes

\begin{equation}
\label{eq:TPF3}
	\langle O_{R(r,s)\alpha}\overline{O}_{T(t,u)\beta} \rangle  =  \delta_{RT}\delta_{rt}\delta_{su}\delta_{\alpha\beta} \frac{(2q)!n!m!2^{q}}{d_{R}\big( (2n)!(2m)! \big)^{2}}\sum\limits_{\tau\in S_{2q}}\mrn{Tr}\big(\hat{P}_{[S]}\Gamma_{R}(\tau)\big)C^{4\nu}_{I}C^{L}_{4\nu}(\tau)^{I}_{L}
\end{equation}
\\
where we also used the fact that 

\begin{equation}
\label{eq:27}
	\bra{[A],[A]\gamma}P_{R\rightarrow(r,s)\gamma\zeta}\ket{[A],[A],\zeta} = 1
\end{equation}
\\
to simplify the two-point function. In (\ref{eq:TPF3}), the $S_{q}[S_{2}]$ in $\hat{P}_{[S]}$ is the stabiliser for $(1,4)(2,3)\cdots$. For reasons that will become clear shortly, we want to change the embedding of the $S_{q}[S_{2}]$. Let $\rho$ be a permutation such that 

\begin{equation}
\label{eq:28}
	C^{4\nu}_{I}(\rho)^{I}_{K} = C^{4\mu}_{K}.
\end{equation}
\\
where $\sigma_{4\mu} = (1,2)(3,4) \cdots (2q-1,2q)$. Defining

\begin{equation}
\label{eq:29}
\tau = \rho^{-1}\psi\rho
\end{equation}
\\ 
equation (\ref{eq:TPF3}) becomes

\begin{equation}
\label{eq:TPF4}
	\langle O_{R(r,s)\alpha}\overline{O}_{T(t,u)\beta} \rangle  = \delta_{RT}\delta_{rt}\delta_{su}\delta_{\alpha\beta}  \frac{(2q)!n!m!2^{q}}{d_{R}\big( (2n)!(2m)! \big)^{2}}\sum\limits_{\psi\in S_{2q}}\mrn{Tr}\big(\hat{P}_{[S]}\Gamma_{R}(\rho^{-1}\psi\rho)\big)C^{4\mu}_{K}C^{L}_{4\mu}(\psi)^{K}_{L}\\
\end{equation}
\\
With this relabelling, we are now contracting indices $1\&2$, $3\&4$ $\cdots (2q-1\&2q)$,

\begin{equation}
\label{eq:30}
	C^{4\mu}_{K}C^{L}_{4\mu}(\psi)^{K}_{L} = (\psi)^{k_{1}k_{1}k_{2}k_{2}\cdots k_{q}k_{q}}_{l_{1}l_{1}l_{2}l_{2}\cdots l_{q}l_{q}}
\end{equation}
\\
Inside the trace, $\Gamma_{R}(\rho)\hat{P}_{[S]}\Gamma_{R}(\rho^{-1}) $ is now a sum over $S_{q}[S_{2}]$ which stabilises the set $(1,2),(3,4)\cdots (2q-1,2q)$. We now have the following for the two-point function

\begin{equation}
	\langle O_{R(r,s)\alpha}\overline{O}_{T(t,u)\beta} \rangle  =  \delta_{RT}\delta_{rt}\delta_{su}\delta_{\alpha\beta} \frac{(2q)!n!m!2^{q}}{d_{R}\big( (2n)!(2m)! \big)^{2}}\sum\limits_{\psi\in S_{2q}}\mrn{Tr}\big(P_{[S]}\Gamma_{R}(\psi)\big)C^{4\mu}_{K}C^{L}_{4\mu}(\psi)^{K}_{L}\\
\end{equation}
\\
Now split the sum over $S_{2q}$ up into a sum over the Brauer algrebra $\mc{B}_{q}$ and a sum over $S_{q}[S_{2}]$ as was done in \cite{SON1}. $\mc{B}_{q}$ is isomorphic to the coset $S_{2q}/S_{q}[S_{2}]$.  

\begin{equation}
\label{eq:TPF5}
 	\langle O_{R(r,s)\alpha}\overline{O}_{T(t,u)\beta} \rangle  = \delta_{RT}\delta_{rt}\delta_{su}\delta_{\alpha\beta}\frac{(2q)!n!m!2^{q}}{d_{R}\big( (2n)!(2m)! \big)^{2}}\sum\limits_{\psi_{1}\in \mc{B}_{q}}\sum\limits_{\psi_{2}\in S_{q}[S_{2}]}\mrn{Tr}\big(P_{[S]}\Gamma_{R}(\psi_{1}\psi_{2})\big)C^{4\mu}_{K}C^{L}_{4\mu}(\psi_{1}\psi_{2})^{K}_{L}
\end{equation}
\\
Now $\psi_{2}$ is an element of the stabiliser of $\sigma_{4\mu}$, which means that

\begin{equation}
\label{eq:31}
	C^{4\mu}_{K}(\psi_{2})^{K}_{I} = C^{4\mu}_{I}
\end{equation}
\\
Then

\begin{equation}
\label{eq:32}
 	\langle O_{R(r,s)\alpha}\overline{O}_{T(t,u)\beta} \rangle  = \delta_{RT}\delta_{rt}\delta_{su}\delta_{\alpha\beta}\frac{(2q)!q!n!m!2^{2q}}{d_{R}\big( (2n)!(2m)! \big)^{2}}\sum\limits_{\psi_{1}\in \mc{B}_{q}}\mrn{Tr}\big(\Gamma_{R}(\psi_{1})P_{[S]}\big)C^{4\mu}_{I}C^{L}_{4\mu}(\psi_{1})^{I}_{L}
\end{equation}
\\
Since $\mc{B}_{q}$ is the set of all coset representatives, and therefore not unique, we again choose the following elements for $\mc{B}_{q}$ \cite{SON1}\footnote{Indeed, this is why we changed the embedding  in the first place - so we could use the permutations in (\ref{eq:CosetReps}) as the set of coset representatives.}. 

\begin{equation}
\label{eq:CosetReps}
		\prod^{q-1}_{j=0}\prod^{2j+1}_{i=1}(i,2j+1)
\end{equation}
\\
It then follows that

\begin{equation}
\label{eq:TPFformula}
	\langle O_{R(r,s)\alpha}\overline{O}_{T(t,u)\beta} \rangle  = \delta_{RT}\delta_{rt}\delta_{su}\delta_{\alpha\beta} \frac{(2q)!q!n!m!2^{2q}}{d_{R}((2n)!(2m)!)^{2}}\bra{[S]}\big(\prod^{q-1}_{j=0}(N+J_{2j+1})\ket{[S]}
\end{equation}
\\
where we wrote $P_{[S]} = \ket{[S]}\!\!\bra{[S]}$. $J_{2j+1}$ is the Jucys-Murphy element in irrep $R$. $J_{2j+1}$ acting on a state in $R$ returns the content of the box labeled $2j+1$. Thus, $N + j_{2j+1}$ gives the weight of that box. Keep in mind that the $\ket{[S]}$ in (\ref{eq:TPFformula}) is symmetric in boxes $(1,2)$ and $(3,4) \cdots$, whereas the $\ket{[S]}$ in the definition of the restricted Schurs is symmetric in boxes $(1,4)$ and $(2,3)$ and so on.

The state $\ket{[S]}$ in (\ref{eq:TPFformula}) only consists of states in $R$ that has boxes $1\&2$ next to each other, and $3\&4$ next to each other $\cdots$ and boxes $2q-1\&2q$ next to each other. As examples, we have

\begin{equation}
\label{eq:33}
	\begin{Young} 1&2&3&4\cr5&6\cr7&8\cr \end{Young} \hspace{20pt} and \hspace{20pt}\begin{Young} 1&2&5&6\cr3&4\cr7&8\cr \end{Young}
\end{equation}
\\
contributing to $\ket{[S]}$ for this particular $R$. In the end, it is only boxes in the odd columns whose weights contribute to the product in (\ref{eq:TPFformula}). By odd columns, we mean the weights of the all the boxes in the 1st column and the weights of all the boxes in the 3rd column and then the 5th column and so on. For example we take the weights of the starred boxes

\begin{equation}
\label{eq:34}
	\begin{Young} *&\cr*&\cr \end{Young},\;\;\;\begin{Young} *&&*&\cr*&&*&\cr \end{Young}, \;\;\;\begin{Young} *&\cr*&\cr*&\cr*&\cr \end{Young}, \;\;\; \begin{Young} *&&*&\cr*&\cr*&\cr \end{Young}
\end{equation}
\\
We then arrive at the following formula for the two-point function

\begin{equation}
\label{eq:Finalformula}
	\langle O_{R(r,s)\alpha} \overline{O}_{T(t,u)\beta}\rangle = \delta_{RT}\delta_{rt}\delta_{su}\delta_{\alpha\beta} \frac{(2q)!q!n!m!2^{2q}}{d_{R}\big( (2n)!(2m)! \big)^{2}}\prod\limits_{i\;\in\; odd\;columns\;in\;R} c_{i}
\end{equation}
\\
We test our formula for a variety of $S_{4}$ and $S_{8}$ examples in appendix 2. Unfortunately, the first multiplicity in the $S_{2n}\times S_{2m}$ irreps we found occurred for the case of 

\begin{equation}
	R = \begin{Young} &&&&&\cr&&&\cr&&&\cr&\cr&\cr&\cr & \cr \end{Young} \hspace{20pt} \mrn{subducing\;2\;copies\;of} \hspace{20pt} (r,s) = ( \begin{Young} &&&\cr&&&\cr&\cr&\cr\cr  \cr \end{Young},  \begin{Young} &&\cr&&\cr\cr \cr\end{Young})
\end{equation}
\\
which is beyond our computational capability.

\section{Discussion}

In this work, we have constructed an exactly orthogonal basis for the $1/4$-BPS sector of free $\mc{N}=4$ super Yang-Mills theory with $SO(2n)$ gauge group. We have shown that the counting of states in this sector exactly matches the number of restricted Schurs the can be defined for $(r,s)$ having even columns and $R$ having even rows. Furthermore, the counting went according to $g(r,s,R)$. This means that there are less operators than for the $U(N)$ case for which the counting went according to $g(r,s,R)^{2}$. We have also presented a variety of simple examples in support of our findings. We then constructed a basis of operators which matched the counting. The basis we constructed also has the nice symmetry property of being invariant under sending

\begin{equation}
\label{eq:35}
	\sigma \rightarrow \eta \sigma \xi \hspace{20pt} \eta \in S_{n}[S_{2}]\times S_{m}[S_{2}] \; \mrn{and}\; \xi \in S_{q}[S_{2}]. 
\end{equation}
\\
Finally we achieved an analytic formula for the two-point function in the form of (\ref{eq:Finalformula}). The restricted Schurs (\ref{eq:RSDef}) are orthogonal and its two-point function is relatively simple to evaluate. We take the product of the weights of the boxes in the odd columns. We also computed a few simple, but non-trivial, examples to check this formula. A next natural step is to study the spectrum of anomalous dimensions of these operators. Using AdS/CFT, we can then study physics of the dual brane objects. We hope to make progress in this direction soon.\\

\emph{Acknowledgements:} I would like to thank Pablo Diaz, Nirina Tahiridimbisoa for many useful discussions. In particular, I'd like to thank Prof. Robert de Mello Koch for all the input and encouragement, and without whom this work would certainly not have been possible. 

\section{Appendix 1. Counting examples}

\subsubsection{1 $Z$ and 1 $Y$ for $SO(4)$}

First consider the case of having only 1 $Z$ and 1 $Y$ for $SO(4)$. The partition function gives a total of 2 operators. We now try to match this number with the number of restricted Schurs that can be defined. There is only 1 possible diagram for $r$,

\begin{equation}
\label{eq:36}
	 \begin{Young} \cr\cr \end{Young} 
\end{equation}
\\
and one possible Young diagram for $s$,

\begin{equation}
\label{eq:37}
	\begin{Young} \cr\cr \end{Young}
\end{equation}
\\
Multiplying $r$ and $s$ together, we find

\begin{equation}
\label{eq:38}
	\begin{Young} \cr\cr \end{Young}\otimes \begin{Young} 1\cr2\cr \end{Young} = 
	\begin{Young}& 1\cr&2\cr \end{Young} \;\otimes \; \begin{Young} \cr\cr1\cr2\cr \end{Young} 
\end{equation}
\\
Using $\delta_{ij}$ and $\epsilon_{ijkl}$, we  can construct only are two operators:

\begin{eqnarray}
\label{eq:39}
	\mc{O}_{1} &=& \mrn{Tr}(ZY)
\label{eq:40}\\
	\mc{O}_{2} &=& \epsilon_{ijkl}Z^{ij}Y^{kl}
\end{eqnarray}
\\
Thus, we have the following correspondence

\begin{eqnarray}
	\mrn{Tr}(ZY) &\leftrightarrow& \begin{Young}& \cr&\cr \end{Young}, (\begin{Young}\cr\cr\end{Young},\begin{Young}\cr\cr\end{Young})\\
	 \epsilon_{ijkl}Z^{ij}Y^{kl} &\leftrightarrow& \begin{Young} \cr\cr\cr\cr \end{Young}, (\begin{Young}\cr\cr\end{Young},\begin{Young}\cr\cr\end{Young})
\end{eqnarray}
\\
where the last operator is known as the Pfaffian \cite{SON1}

\subsubsection{2 $Z$'s and 2 $Y$'s for $SO(4)$} 

For 2 $Z$'s and 2 $Y$'s for $SO(4)$, the partition function gives a total of 7 operators. We thus expect 7 different restricted Schur labels. For $r$ we can have

\begin{equation}
\label{eq:41}
	 \begin{Young}& \cr&\cr \end{Young}, \hspace{20pt}  \begin{Young}\cr\cr\cr\cr \end{Young}
\end{equation}
\\
For $s$ we can have

\begin{equation}
\label{eq:42}
	 \begin{Young}& \cr&\cr \end{Young}, \hspace{20pt}  \begin{Young}\cr\cr\cr\cr \end{Young}
\end{equation}
\\
Multiplying $r$ and $s$ together and only taking the results with even rows, we get

\begin{eqnarray}
	\begin{Young} &\cr&\cr \end{Young} \otimes \begin{Young} 1&1\cr2&2\cr \end{Young} &\mrn{gives}& 
	\begin{Young} &&1&1\cr&&2&2\cr \end{Young}\;,\; \begin{Young} &&1&1\cr&\cr2&2\cr \end{Young} \;,\; \begin{Young} &&1\cr&&2\cr1\cr2\cr \end{Young}\;
	\;,\;\begin{Young} &\cr&\cr 1&1 \cr2 &2\cr \end{Young}
\end{eqnarray}
\\
Next

\begin{eqnarray}
\begin{Young} &\cr&\cr \end{Young} \otimes \begin{Young} 1\cr2\cr3\cr4\cr \end{Young} &\mrn{gives}& \begin{Young} &&1\cr&&2\cr3\cr4\cr \end{Young}
\end{eqnarray}
\\
Then

\begin{eqnarray}
  \begin{Young} \cr\cr\cr\cr \end{Young} \otimes \begin{Young} 1&1\cr2&2\cr \end{Young} &\mrn{gives}& \begin{Young} &1&1\cr&2&2\cr\cr\cr \end{Young}
\end{eqnarray}
\\
Finally

\begin{eqnarray}
  \begin{Young} \cr\cr\cr\cr \end{Young} \otimes  \begin{Young} 1\cr2\cr3\cr4\cr \end{Young}&\mrn{gives}& \begin{Young} &1\cr&2\cr&3 \cr &4\cr \end{Young}
\end{eqnarray}
\\
Counting all the Young diagrams we get 7 with the correct labelling. We also check that we find 7 operators using $\delta_{ij}$ and $\epsilon_{ijkl}$. They are listed below

\begin{eqnarray}
	\mc{O}_{1}&=& \mrn{Tr}(Z^{2}Y^{2}) \nonumber \\
	\mc{O}_{2}&=& \mrn{Tr}(ZYZY) \nonumber \\
	\mc{O}_{3}&=& \mrn{Tr}(Z^{2})\mrn{Tr}(Y^{2})\nonumber  \\
	\mc{O}_{4}&=& \mrn{Tr}(ZY)^{2} \nonumber \\
	\mc{O}_{5} &=& \epsilon_{ijkl}Z^{ij}Z^{kl}\mrn{Tr}(Y^{2}) \nonumber \\
	\mc{O}_{6} &=& \epsilon_{ijkl}Z^{ij}Y^{kl}\mrn{Tr}(ZY) \nonumber \\
	\mc{O}_{7} &=& \epsilon_{ijkl}Y^{ij}Y^{kl}\mrn{Tr}(Z^{2})
\end{eqnarray}
\\
\subsubsection{3 $Z$'s and 2 $Y$'s for $SO(6)$}

Here we do not expect multitrace operators. This is because there is an odd number of fields; we will always end up with a $\mrn{Tr}(Z)$ or $\mrn{Tr}(Y)$ which vanishes. We can, however, obtain operators built using the $\epsilon_{ijklmn}$. The partition function gives a total of 4 operators. We thus expect 4 restricted Schurs and each of the should have the Pfaffian as a factor. For $r$, the possible Young diagrams are

\begin{equation}
\label{eq:43}
	 \begin{Young} &&\cr&&\cr \end{Young}, \hspace{20pt}  \begin{Young} &\cr&\cr\cr\cr \end{Young}, \hspace{20pt}  \begin{Young} \cr\cr\cr\cr\cr\cr \end{Young}
\end{equation}
\\
and for $s$, the possible Young diagrams are

\begin{equation}
\label{eq:44}
	 \begin{Young}& \cr&\cr \end{Young}, \hspace{20pt}  \begin{Young}\cr\cr\cr\cr \end{Young}
\end{equation}
\\
Multiplying each $r$ with each $s$, and taking only the diagrams which contribute, we find

\begin{eqnarray}
	 \begin{Young} &&\cr&&\cr \end{Young} \otimes \begin{Young}1&1 \cr2&2\cr \end{Young} &\mrn{gives}& 0 \\
	 && \nonumber \\
	  \begin{Young} &\cr&\cr\cr\cr \end{Young} \otimes  \begin{Young}1&1 \cr2&2\cr \end{Young} &\mrn{gives}&  \begin{Young} &&1\cr&&2\cr\cr\cr1\cr2\cr \end{Young} \\
	  && \nonumber \\
	   \begin{Young} \cr\cr\cr\cr\cr\cr \end{Young} \otimes \begin{Young}1&1 \cr2&2\cr \end{Young} &\mrn{gives}& \begin{Young} &1&1\cr&2&2\cr\cr\cr\cr\cr \end{Young}\\
	   &&\nonumber \\
	    \begin{Young} &&\cr&&\cr \end{Young} \otimes \begin{Young}1\cr2\cr3\cr4\cr \end{Young} &\mrn{gives}&  \begin{Young} &&\cr&&\cr1\cr 2\cr3\cr4\cr \end{Young}\\
	    && \nonumber \\
	     \begin{Young} &\cr&\cr\cr\cr \end{Young} \otimes  \begin{Young}1\cr2\cr3\cr4\cr \end{Young} &\mrn{gives}&  \begin{Young} &&1\cr&&2\cr\cr\cr3\cr4\cr \end{Young}
\end{eqnarray}
\\
\begin{eqnarray}
	    && \nonumber \\
	     	\begin{Young} \cr\cr\cr\cr\cr\cr \end{Young}\otimes \begin{Young}1\cr2\cr3\cr4\cr \end{Young} &\mrn{gives}& 0
\end{eqnarray}
\\
We indeed obtain 4 restricted Schurs as expected.

\subsubsection{4 $Z$'s and $2$Y's for SO(8)}

The partition function gives a total of 13 operators. The possible diagrams for $r$ are

\begin{equation}
\label{eq:45}
	\begin{Young} &&&\cr&&&\cr \end{Young}, \hspace{20pt} \begin{Young} &\cr&\cr&\cr&\cr \end{Young} , \hspace{20pt} \begin{Young} &&\cr&&\cr\cr \cr \end{Young} , \hspace{20pt} \begin{Young} &\cr&\cr\cr \cr\cr\cr \end{Young}, \hspace{20pt} \begin{Young} \cr\cr\cr\cr\cr\cr\cr\cr \end{Young} 
\end{equation}
\\
and for $s$ 

\begin{equation}
\label{eq:46}
	 \begin{Young}& \cr&\cr \end{Young}, \hspace{20pt}  \begin{Young}\cr\cr\cr\cr \end{Young}
\end{equation}
\\
Multiplying each $r$ with each $s$ and taking only the diagrams which contribute to the counting, we find

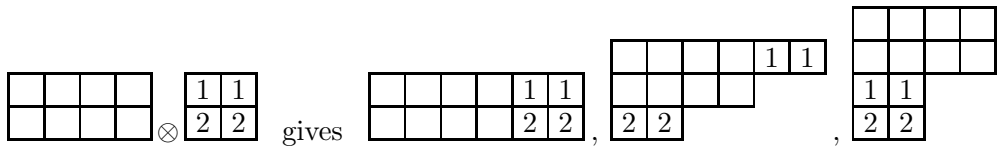
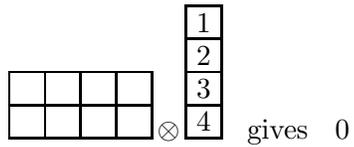
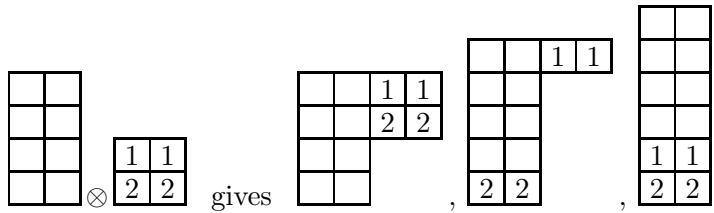
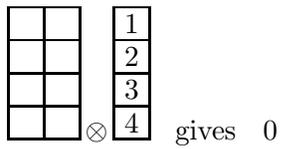
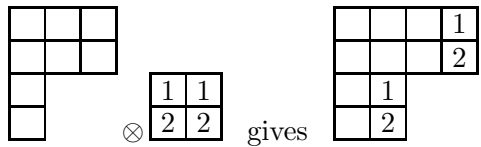
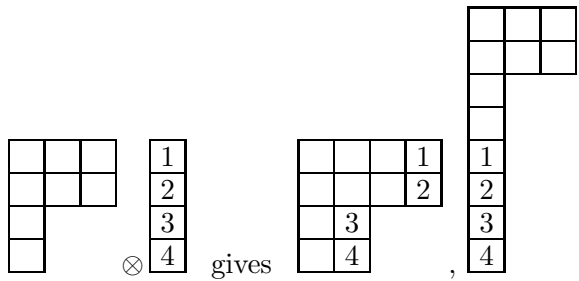
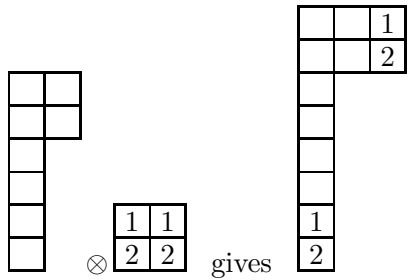
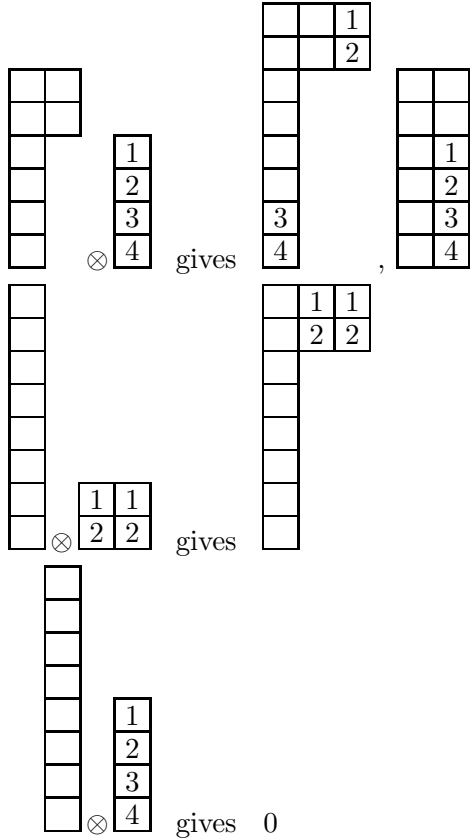
\begin{eqnarray}
	\begin{Young} &&&\cr&&&\cr \end{Young} \otimes  \begin{Young}1&1 \cr2&2\cr \end{Young} &\mrn{gives}& \begin{Young} &&&&1&1\cr&&&&2&2\cr \end{Young} \;,\; \begin{Young} &&&&1&1\cr&&&\cr2&2\cr \end{Young}\;,\;\begin{Young} &&&\cr&&&\cr 1&1\cr2&2\cr\end{Young} \nonumber  \\
	\begin{Young} &&&\cr&&&\cr \end{Young} \otimes  \begin{Young}1\cr2\cr3\cr4\cr \end{Young} &\mrn{gives}& 0 \nonumber \\
	 \begin{Young} &\cr&\cr&\cr&\cr \end{Young} \otimes \begin{Young}1&1 \cr2&2\cr \end{Young} &\mrn{gives}&  \begin{Young} &&1&1\cr&&2&2\cr&\cr&\cr \end{Young}\;,\; \begin{Young} &&1&1\cr&\cr&\cr&\cr2&2\cr \end{Young}\;,\;\begin{Young} &\cr&\cr&\cr&\cr 1&1\cr2 &2\cr\end{Young} \nonumber \\
	  \begin{Young} &\cr&\cr&\cr&\cr \end{Young} \otimes   \begin{Young}1\cr2\cr3\cr4\cr \end{Young} &\mrn{gives}& 0\\
	   \begin{Young}& &\cr&&\cr\cr \cr \end{Young} \otimes  \begin{Young}1&1 \cr2&2\cr \end{Young} &\mrn{gives}&   \begin{Young}&&&1\cr&&&2\cr&1\cr&2 \cr \end{Young} \nonumber \\
	    \begin{Young}&& \cr&& \cr\cr\cr \end{Young} \otimes  \begin{Young}1\cr2\cr3\cr4\cr \end{Young} &\mrn{gives}& \begin{Young}&&&1 \cr&&&2 \cr&3\cr&4\cr \end{Young} \;,\; \begin{Young}&& \cr&& \cr\cr\cr1\cr2\cr3\cr4 \cr\end{Young} \nonumber \\
	    	 \begin{Young}& \cr& \cr\cr\cr\cr\cr \end{Young} \otimes  \begin{Young}1&1 \cr2&2\cr \end{Young} &\mrn{gives}& \begin{Young}&&1 \cr&&2 \cr\cr\cr\cr\cr1\cr2\cr \end{Young}\nonumber
\end{eqnarray}
\\
\begin{eqnarray*}
		\begin{Young}& \cr& \cr\cr\cr\cr\cr \end{Young} \otimes  \begin{Young}1\cr2\cr3\cr4\cr \end{Young} &\mrn{gives}&
	 \begin{Young}&&1 \cr&&2 \cr\cr\cr\cr\cr3\cr4\cr \end{Young}\;,\;\begin{Young}& \cr& \cr&1\cr&2\cr&3\cr&4\cr \end{Young}	\\
	 \begin{Young} \cr\cr\cr\cr\cr\cr\cr\cr \end{Young} \otimes \begin{Young}1&1 \cr2&2\cr \end{Young} &\mrn{gives}& \begin{Young} &1&1\cr&2&2\cr\cr\cr\cr\cr\cr\cr \end{Young}\\
	 	\begin{Young} \cr\cr\cr\cr\cr\cr\cr\cr \end{Young} \otimes  \begin{Young}1\cr2\cr3\cr4\cr \end{Young}  &\mrn{gives}& 0
\end{eqnarray*}
\\
Adding up all the Young diagram labels, we indeed get 13 operators. The reader is invited to check that for $SO(4)$, there are 12 operators and for $SO(6)$, there are 9 operators.

\subsubsection{3 $Z$'s and 3 $Y$'s for $SO(8)$}

As a final example, we consider 3 $Z$'s and 3 $Y$'s for $SO(8)$. The partition function gives a total of 14 operators. For $r$, we have

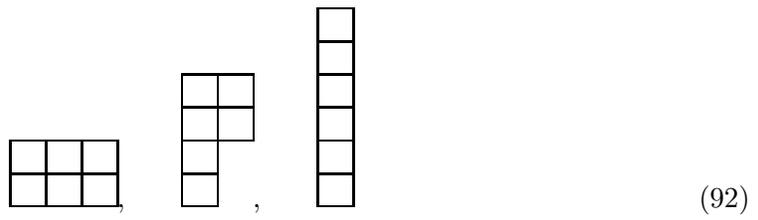
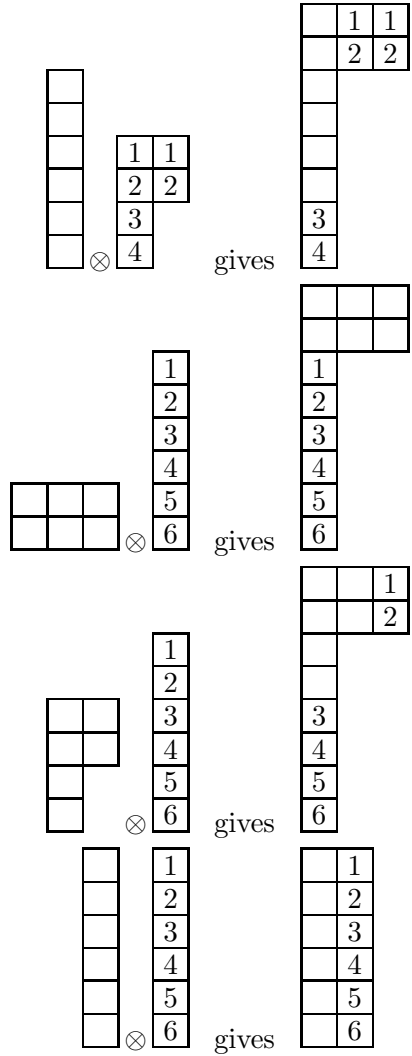
\begin{equation}
\label{eq:47}
	 \begin{Young} &&\cr&&\cr \end{Young}, \hspace{20pt}  \begin{Young} &\cr&\cr\cr\cr \end{Young}, \hspace{20pt}  \begin{Young} \cr\cr\cr\cr\cr\cr \end{Young}
\end{equation}
\\
and for $s$

\begin{equation}
\label{eq:48}
		 \begin{Young} &&\cr&&\cr \end{Young}, \hspace{20pt}  \begin{Young} &\cr&\cr\cr\cr \end{Young}, \hspace{20pt}  \begin{Young} \cr\cr\cr\cr\cr\cr \end{Young}
\end{equation}
\\
Multiplying, we find

\begin{eqnarray}
	 \begin{Young} &&\cr&&\cr \end{Young} \otimes  \begin{Young} 1&1&1\cr2&2&2\cr \end{Young} &\mrn{gives}&  \begin{Young} &&&1&1&1\cr&&&2&2&2\cr \end{Young} \;,\;  \begin{Young} &&&1&1&1\cr&&&2\cr2&2\cr \end{Young},\; \begin{Young} &&&1\cr&&&2\cr1&1\cr2&2\cr \end{Young} \nonumber \\
	 	  \begin{Young} &\cr&\cr\cr\cr \end{Young}\otimes  \begin{Young} 1&1&1\cr2&2&2\cr \end{Young} &\mrn{gives}& \begin{Young} &&1&1\cr&&2&2\cr&1\cr&2\cr \end{Young} \nonumber \\
		  \begin{Young} \cr\cr\cr\cr\cr\cr \end{Young} \otimes  \begin{Young} 1&1&1\cr2&2&2\cr \end{Young} &\mrn{gives}& \begin{Young} &1&1 \cr&2&2\cr\cr\cr\cr\cr1\cr2\cr \end{Young} \nonumber \\
		   \begin{Young} &&\cr&&\cr \end{Young} \otimes \begin{Young} 1&1\cr2&2\cr.\cr4\cr \end{Young} &\mrn{gives}& \begin{Young} &&&1\cr&&&2\cr1&3\cr2&4\cr \end{Young}\\
		   \begin{Young} &\cr&\cr\cr\cr \end{Young}\otimes  \begin{Young} 1&1\cr2&2\cr3\cr4\cr \end{Young} &\mrn{gives}&  \begin{Young}&&1&1\cr&&2&2\cr&3\cr&4\cr \end{Young}\;,\;  \begin{Young} &&1&1\cr&\cr&2\cr&3\cr2&4\cr \end{Young}\;,\; \begin{Young} &&1\cr&&2\cr\cr\cr1\cr2\cr3\cr4\cr \end{Young} \;,\; \begin{Young} &\cr&\cr&1\cr&2\cr1&3\cr2&4\cr \end{Young} \nonumber
\end{eqnarray}
\\
\begin{eqnarray*}
	\begin{Young} \cr\cr\cr\cr\cr\cr \end{Young}\otimes  \begin{Young} 1&1\cr2&2\cr3\cr4\cr \end{Young} &\mrn{gives}& \begin{Young} &1&1\cr&2&2\cr\cr\cr\cr\cr3\cr4\cr \end{Young}\\
	\begin{Young} &&\cr&&\cr \end{Young}\otimes \begin{Young} 1\cr2\cr3\cr4\cr5\cr6\cr \end{Young} &\mrn{gives}& \begin{Young} &&\cr&&\cr1\cr2\cr3\cr4\cr5\cr6\cr \end{Young} \\
	 \begin{Young} &\cr&\cr\cr\cr \end{Young}\otimes \begin{Young} 1\cr2\cr3\cr4\cr5\cr6\cr \end{Young} &\mrn{gives}& \begin{Young} &&1\cr&&2\cr\cr\cr3\cr4\cr5\cr6\cr \end{Young}\\
	 \begin{Young} \cr\cr\cr\cr\cr\cr \end{Young}\otimes \begin{Young} 1\cr2\cr3\cr4\cr5\cr6\cr \end{Young} &\mrn{gives}&  \begin{Young} &1\cr&2\cr&3\cr&4\cr&5\cr&6\cr \end{Young}
\end{eqnarray*}
\\
Adding up all the Young diagram labels, we get our expected 14 restricted Schur polynomials. Again the reader is invited to check that for $SO(6)$ there are 10 restricted Schurs and for $SO(4)$, there are 12 restricted Schurs.

\section{Appendix 2. Two-point function examples}

In the following examples, we denote the eigenvector of $\hat{P}_{[S]}$ by $\hat{\ket{[S]}}$.

\subsection{$q=2$}

For $q=2$, we have

\begin{equation}
\label{eq:50}
	R = \begin{Young}&\cr&\cr\end{Young}, \hspace{20pt} (r,s) = (\begin{Young}\cr\cr\end{Young},\begin{Young}\cr\cr\end{Young})
\end{equation}
\\
Firstly

\begin{equation}
\label{eq:51}
	\ket{[\hat{S}]} = \frac{1}{2}\,\begin{Young}4&3\cr2&1\cr\end{Young} + \frac{\sqrt{3}}{2}\,\begin{Young}4&2\cr3&1\cr\end{Young}, \hspace{20pt} \ket{[A],[A]} = \begin{Young}4&2\cr3&1\cr\end{Young}
\end{equation}
\\
The restricted Schur is

\begin{equation}
\label{eq:52}
	O_{R(r,s)} = 2\sqrt{3}\mrn{Tr}(ZY)
\end{equation}
\\
Evaluating the Wick contractions, its two point function is

\begin{equation}
\label{eq:53}
	\langle O_{R(r,s)}\overline{O}_{T(t,u)} \rangle = 12 \langle \mrn{Tr}(ZY) \mrn{Tr}(\overline{Z}\overline{Y}) \rangle = 24N(N-1)
\end{equation}
\\
Using (\ref{eq:TPFformula}), with $q=2, n=m=1$ and $d_{R}=2$, we get

\begin{equation}
\label{eq:54}
\langle O_{R(r,s)}\overline{O}_{T(t,u)} \rangle = 24N(N-1).
\end{equation}
\\
\subsection{$q=4$}

First consider

\begin{equation}
\label{eq:55}
	R = \begin{Young} &&&\cr&&&\cr \end{Young}, \hspace{20pt} (r,s) = (\begin{Young} &\cr&\cr \end{Young},\begin{Young} &\cr&\cr \end{Young})
\end{equation}
\\
We caluculate

\begin{equation}
\label{eq:ShatforR44}
	\ket{[\hat{S}]} = \frac{\sqrt{5}}{4} \begin{Young} 8&6&4&2\cr7&5&3&1\cr \end{Young} + \sqrt{\frac{5}{48}} \begin{Young} 8&7&4&2\cr6&5&3&1\cr \end{Young} + \frac{\sqrt{5}}{12} \begin{Young} 8&7&4&3\cr6&5&2&1\cr \end{Young} + \frac{2}{3} \begin{Young} 8&7&6&5\cr4&3&2&1\cr \end{Young} + \sqrt{\frac{5}{48}}\begin{Young} 8&6&4&3\cr7&5&2&1\cr \end{Young}
\end{equation}
\\
and 

\begin{equation}
\label{eq:AAvecforR44}
	\ket{[A],[A]} = \begin{Young} 8&6&4&2\cr7&5&3&1\cr \end{Young}
\end{equation}
\\
The operator is

\begin{equation}
\label{eq:56}
	O_{R(r,s)} = \frac{2\sqrt{5}}{3}\Big( 2\mrn{Tr}(ZY)^{2} + 2\mrn{Tr}(ZYZY) + \mrn{Tr}(Z^{2})\mrn{Tr}(Y^{2}) + 4\mrn{Tr}(Z^{2}Y^{2}) \Big)
\end{equation}
\\
and, after evaluating all the Wick contractions, its two-point function is

\begin{equation}
\label{eq:57}
	\langle O_{R(r,s)}\overline{O}_{R(r,s)} \rangle = \frac{640}{3}N(N-1)(N+1)(N+2)
\end{equation}
\\
Equation (\ref{eq:TPFformula}), for $q=4, n=m=2, d_{R} = 14$ and taking only the weights of the odd columns, gives

\begin{equation}
\label{eq:58}
	\langle O_{R(r,s)}\overline{O}_{R(r,s)} \rangle = \frac{640}{3}N(N-1)(N+1)(N+2)
\end{equation}
\\
Next consider

\begin{equation}
\label{eq:59}
	R = \begin{Young} &\cr&\cr&\cr&\cr \end{Young}, \hspace{20pt} (r,s) = (\begin{Young} &\cr&\cr \end{Young},\begin{Young} &\cr&\cr \end{Young})
\end{equation}
\\
The state $\ket{[\hat{S}]}$ was found to be

\begin{equation}
\label{eq:60}
	\ket{[\hat{S}]} = \frac{3}{4} \begin{Young}8 &6\cr7&5\cr4&2\cr3&1\cr \end{Young} + \frac{\sqrt{3}}{4} \begin{Young}8 &7\cr6&5\cr4&2\cr3&1\cr \end{Young} + \frac{1}{4} \begin{Young}8 &7\cr6&5\cr4&3\cr2&1\cr \end{Young} + \frac{\sqrt{3}}{4} \begin{Young} 8&6\cr7&5\cr4&3\cr2&1\cr \end{Young} 
\end{equation}
\\
and 

\begin{equation}
\label{eq:61}
	\ket{[A],[A]} = \begin{Young}8 &6\cr7&5\cr4&2\cr3&1\cr \end{Young}
\end{equation}
\\
The operator is

\begin{equation}
\label{eq:62}
	O_{R(r,s)} = \frac{2}{3}\Big( 2\mrn{Tr}(ZY)^{2} + 2\mrn{Tr}(ZYZY) + 3\mrn{Tr}(Z^{2})\mrn{Tr}(Y^{2}) - 12\mrn{Tr}(Z^{2}Y^{2}) \Big)
\end{equation}
\\
and its two-point function is

\begin{equation}
\label{eq:63}
	\langle O_{R(r,s)}\overline{O}_{R(r,s)} \rangle = \frac{640}{3}N(N-1)(N-2)(N-3)
\end{equation}
\\
Equation (\ref{eq:TPFformula}), for $q=4, n=m=2, d_{R} = 14$ and taking the weights from the odd columns, gives

\begin{equation}
\label{eq:64}
	\langle O_{R(r,s)}\overline{O}_{R(r,s)} \rangle = \frac{640}{3}N(N-1)(N-2)(N-3)
\end{equation}
\\
Next consider

\begin{equation}
\label{eq:65}
	R = \begin{Young} &\cr&\cr&\cr&\cr \end{Young}, \hspace{20pt} (r,s) = (\begin{Young} \cr\cr\cr\cr \end{Young},\begin{Young} \cr\cr\cr\cr \end{Young})
\end{equation}
\\
We find

\begin{equation}
\label{eq:ShatforR2222}
	\ket{[\hat{S}]} = \frac{3}{4} \begin{Young}8 &6\cr7&5\cr4&2\cr3&1\cr \end{Young} + \frac{\sqrt{3}}{4} \begin{Young}8 &7\cr6&5\cr4&2\cr3&1\cr \end{Young} + \frac{1}{4} \begin{Young}8 &7\cr6&5\cr4&3\cr2&1\cr \end{Young} + \frac{\sqrt{3}}{4} \begin{Young} 8&6\cr7&5\cr4&3\cr2&1\cr \end{Young} 
\end{equation}
\\
and 

\begin{equation}
\label{eq:66}
	\ket{[A],[A]} = \begin{Young}8 &4\cr7&3\cr6&2\cr5&1\cr \end{Young}
\end{equation}
\\
The operator is

\begin{equation}
\label{eq:67}
	O_{R(r,s)} = \frac{4\sqrt{5}}{3}\Big( \mrn{Tr}(ZY)^{2} - 2\mrn{Tr}(ZYZY) \Big)
\end{equation}
\\
and its two-point function is

\begin{equation}
\label{eq:68}
	\langle O_{R(r,s)}\overline{O}_{R(r,s)} \rangle = \frac{640}{3}N(N-1)(N-2)(N-3)
\end{equation}
\\
Equation (\ref{eq:TPFformula}) gives 

\begin{equation}
\label{eq:69}
	\langle O_{R(r,s)}\overline{O}_{R(r,s)} \rangle = \frac{640}{3}N(N-1)(N-2)(N-3)
\end{equation}
\\
For the final example for which $n=m=2$, consider

\begin{equation}
\label{eq:6911}
	R = \begin{Young} &&&\cr&\cr&\cr \end{Young}, \hspace{20pt} (r,s) = (\begin{Young} &\cr&\cr \end{Young},\begin{Young} &\cr&\cr \end{Young})
\end{equation}
\\
We find the following for $\ket{[\hat{S}]}$ and $\ket{[A],[A]}$

\begin{eqnarray}
	\ket{[\hat{S}]} &=& \frac{1}{4\sqrt{6}}\begin{Young} 8&7&2&1\cr6&5\cr4&3\cr \end{Young} + \frac{1}{4\sqrt{2}}\begin{Young} 8&6&2&1\cr7&5\cr4&3\cr \end{Young} + \frac{1}{8\sqrt{3}}\begin{Young} 8&7&3&1\cr6&5\cr4&2\cr \end{Young} + \frac{1}{8}\begin{Young} 8&6&3&1\cr7&5\cr4&2\cr \end{Young} + \frac{\sqrt{5}}{8}\begin{Young} 8&7&4&1\cr6&5\cr3&2\cr \end{Young} \nonumber \\
	&&+\; \frac{\sqrt{15}}{8}\begin{Young} 8&6&4&1\cr7&5\cr3&2\cr \end{Young} - \frac{1}{12\sqrt{2}}\begin{Young} 8&7&4&3\cr6&5\cr2&1\cr \end{Young} - \frac{1}{4\sqrt{6}}\begin{Young} 8&6&4&3\cr7&5\cr2&1\cr \end{Young} - \frac{1}{8\sqrt{3}}\begin{Young} 8&7&4&2\cr6&5\cr3&1\cr \end{Young} - \frac{1}{8}\begin{Young} 8&6&4&2\cr7&5\cr3&1\cr \end{Young} \nonumber \\
	&&+\;\frac{\sqrt{5}}{8}\begin{Young} 8&7&3&2\cr6&5\cr4&1\cr \end{Young} + \frac{\sqrt{15}}{8}\begin{Young} 8&6&3&2\cr7&5\cr4&1\cr \end{Young} + \frac{1}{6}\sqrt{\frac{5}{2}}\begin{Young} 8&7&6&5\cr4&3\cr2&1\cr \end{Young} + \frac{1}{2}\sqrt{\frac{5}{6}}\begin{Young} 8&7&6&5\cr4&2\cr3&1\cr \end{Young}
\end{eqnarray}
\\
and 

\begin{equation}
\label{eq:70}
	\ket{[A],[A]} = -\frac{3}{8}\begin{Young} 8&6&3&1\cr7&5\cr4&2\cr \end{Young} + \frac{\sqrt{15}}{8}\begin{Young} 8&6&4&1\cr7&5\cr3&2\cr \end{Young} - \frac{5}{8}\begin{Young} 8&6&4&2\cr7&5\cr3&1\cr \end{Young} + \frac{\sqrt{15}}{8}\begin{Young} 8&6&3&2\cr7&5\cr4&1\cr\end{Young} 
\end{equation}
\\
The operator is

\begin{equation}
\label{eq:71}
	O_{R(r,s)} = \frac{4}{3}\Big( -\mrn{Tr}(ZY)^{2} - \mrn{Tr}(ZYZY) + \mrn{Tr}(Z^{2})\mrn{Tr}(Y^{2}) + \mrn{Tr}(Z^{2}Y^{2}) \Big)
\end{equation}
\\
and its two-point function, after evaluating all the Wick contractions, is

\begin{equation}
\label{eq:72}
	\langle O_{R(r,s)}\overline{O}_{R(r,s)} \rangle = \frac{160}{3}N(N-1)(N-2)(N+2)
\end{equation}
\\
Equation (\ref{eq:TPFformula}), for $q=4, n=m=2, d_{R} = 56$ and odd columns, gives

\begin{equation}
\label{eq:73}
	\langle O_{R(r,s)}\overline{O}_{R(r,s)} \rangle = \frac{160}{3}N(N-1)(N-2)(N+2)
\end{equation}
\\
Now let's try two examples where $n=3, m=1$, i.e., the number of $Z$'s is not the equal to the number of $Y$'s. Consider

\begin{equation}
\label{eq:74}
	R = \begin{Young} &&&\cr&&&\cr \end{Young}, \hspace{20pt} (r,s) = (\begin{Young} &&\cr&&\cr \end{Young},\begin{Young} \cr\cr \end{Young})
\end{equation}
\\
States $\hat{\ket{[S]}}$ and $\ket{[A],[A]}$ are still the same as in (\ref{eq:ShatforR44}) and (\ref{eq:AAvecforR44}). The operator is

\begin{equation}
\label{eq:75}
	O_{R(r,s)} = \frac{4}{\sqrt{5}}\Big( \mrn{Tr}(ZY)\mrn{Tr}(Z^{2}) + 2 \mrn{Tr}(Z^{3}Y) \Big)
\end{equation}
\\
and its two-point function is

\begin{equation}
\label{eq:76}
	\langle O_{R(r,s)}\overline{O}_{R(r,s)} \rangle =  \frac{256}{5}N(N-1)(N+1)(N+2).
\end{equation}
\\
Formula (\ref{eq:TPFformula}) reproduces exactly this result. As the final example, we consider

\begin{equation}
\label{eq:77}
	R = \begin{Young} &\cr&\cr&\cr&\cr \end{Young}, \hspace{20pt} (r,s) = ( \begin{Young} &\cr&\cr\cr\cr \end{Young}, \begin{Young} \cr\cr \end{Young})
\end{equation}
\\
The $\hat{\ket{[S]}}$ is the same as in (\ref{eq:ShatforR2222}). However

\begin{equation}
\label{eq:78}
	\ket{[A],[A]} = \frac{\sqrt{5}}{3}\begin{Young} 8&4\cr7&3\cr6&2\cr5&1\cr\end{Young} + \frac{2}{3}\begin{Young} 8&6\cr7&5\cr4&2\cr3&1\cr\end{Young}
\end{equation}
\\
The operator is

\begin{equation}
\label{eq:79}
	O_{R(r,s)} = \frac{4}{\sqrt{5}}\Big( \mrn{Tr}(ZY)\mrn{Tr}(Z^{2}) - 2 \mrn{Tr}(Z^{3}Y) \Big)
\end{equation}
\\
The two-point function is

\begin{equation}
\label{eq:80}
	\langle O_{R(r,s)}\overline{O}_{R(r,s)} \rangle = \frac{256}{5}N(N-1)(N-2)(N-3).
\end{equation}
\\
which, again, is exactly reproduced by (\ref{eq:TPFformula}).



\bibliographystyle{unsrt}

\bibliography{Refs}

\end{document}